# Title

Twisted Radio Waves and Twisted Thermodynamics

## Authors and Affiliations

L.B. Kish
corresponding author
Solid State Electronics, Photonics, and Nanoengineering Division, Department of Electrical and Computer Engineering, Texas A&M University, College Station, TX 77843-3128, USA

Robert D. Nevels
Electro-Magnetics and Microwaves Division, Department of Electrical and Computer Engineering, Texas A&M University, College Station, Texas, 77843-3128, USA

## Abstract

We present and analyze a gedanken experiment and show that the assumption that an antenna operating at a single frequency can transmit more than two independent information channels to the far field violates the Second Law of Thermodynamics. Transmission of a large number of channels, each associated with an angular momenta 'twisted wave' mode, to the far field in free space is therefore not possible.

## Introduction

Recently it has been claimed in scientific literature that it is possible to generate radio waves, at a single frequency, with more spatially orthogonal modes, "orbital modes", than the usual two polarization modes [1-4]. An experimental demonstration with $N = 2$ modes purporting to confirm the twisted wave concept has been carried out and published [2]. Such radio waves would have angular momenta, also referred to as orbital angular momenta, in a way similar to the orthogonal ($l$) wave modes of electrons that exist at the same frequency and belong to the same main quantum number ($n$). Communication utilizing such independent/orthogonal modes would expand the available frequency band by a factor given by the number of additional spatially orthogonal modes. Because the information channel capacity of radio waves scales linearly with the number of spatially orthogonal modes $N$, in the case of fixed bandwidth and signal-to-noise ratio if $N$ can be more than two or even infinite as claimed [1], wireless communication would be revolutionized.

It is important to note that recently two independent groups published papers [5,6] concluding that the proposed twisted wave schemes are a special case of the traditional *multiple-input-multiple-output* (MIMO) technique and are thus not conceptually new.

Furthermore [5] points out that in the far field the twisted wave scheme does not provide any increase in information channel capacity. Paper [6] shows that the experiments [2] have not been performed in "far-enough-field" conditions. A true far field wireless experiment would show further losses and other deficiencies in individual twisted wave modes [5].

## Discussion and Results

In this paper we address a fundamental physics question: Can modes with non-zero angular momenta representing extra, beyond $N = 2$, independent communication channels be radiated to the far field and selectively picked up by a proper antenna, which is insensitive to standard plane wave modes? If the polarization is circular—a common situation in wireless technology—one has $N = 2$ with plane waves in the two polarization modes phase-shifted by 90°. Thus it is clear that up to $N = 2$ orthogonal plane wave polarization modes can exist in the far field and the circularly polarized mode carries angular momentum. Yet to date whether a greater number of angular momentum modes can exist at the same frequency and carrying independent signals in the far field has not been shown to violate fundamental physical principles.

It should first be noted that based on physical principles the assumption that there can be more than the two *far-field* polarization modes is counter-intuitive. In the atom, the existence of waves with different angular momenta at the same energy originates from the potential and the ensuing localized nature of the waves. A charge revolving in a Coulomb potential field will have an infinite number of different classical physical paths with the same energy, and Bohr-Sommerfeld quantization will select a finite number of states that are allowed within quantum theory. But, in stark contrast no such state components exist for free electron waves. In light of this intuitive argument, the existence of spatially orthogonal modes for electromagnetic waves is fine for photons propagating under spatially confined conditions such as in wave guides and optical fibers [7,8], or in the immediate surroundings of a black hole [9]. We reiterate that the existing experimental radio wave demonstrations [2] hold only for $N = 2$.

Rather than analyzing the theoretical treatments for errors, we use another approach to prove that the hypothesis that independent communication channels based on orbital modes can be selectively picked up by a proper antenna that is insensitive to standard plane wave modes violates the Second Law of Thermodynamics, which states that it is impossible to construct a perpetual motion machine of the second kind. First let us specify the necessary conditions that are essential for the utilization of the *M*-th orbital mode as a parallel independent information channel:

*i*) A selective antenna must exist that is able to radiate in the *M*-th orbital mode.

*ii*) The same antenna should selectively pick up a signal from an electromagnetic wave only at the *M*-th orbital mode while discarding all the other orbital and non-orbital mode components in that signal.

According to Planck's Law [10], the a black-body (with unity emissivity) radiates in each polarization with a power spectral intensity

$$I(f) = \frac{4\pi h f^3}{c^2} \frac{1}{e^{hf/kT} - 1} \quad , \tag{1}$$

where $f$ is frequency, $h = 6.626 * 10^{-34}$ Js is Planck's constant, $k = 1.381 * 10^{-23}$ JK$^{-1}$ is Boltzmann's constant and $T$ is absolute temperature. This means that a unit surface area of the black-body emits, in each polarization, the power

$$P(f, \Delta f) = I(f)\Delta f = \frac{4\pi h f^3}{c^2} \frac{\Delta f}{e^{hf/kT} - 1} \tag{2}$$

within an infinitesimally small frequency band $\Delta f$ around $f$. Thus the total radiated power from a unit area is

$$NP(f, \Delta f) = NI(f)\Delta f = N \frac{4\pi h f^3}{c^2} \frac{\Delta f}{e^{hf/kT} - 1} \quad , \tag{3}$$

where $N=2$ is the number of orthogonal polarization modes. Thus the Planck formula [10], is:

$$P(f, \Delta f) = \frac{8\pi h f^3}{c^2} \frac{\Delta f}{e^{hf/kT} - 1} \quad . \tag{4}$$

Inspired by Nyquist's treatment of Johnson noise [11], we now devise the following gedanken experiment, see Figure 1: A large box (much larger than the wavelengths considered) is located in a thermal reservoir of temperature T. We assume that its internal walls are ideally black. Furthermore an isolated resistor (with radiation screening and thermal isolation) and a "twisted-wave" antenna tuned to the *M*-th orbital mode in a bandwidth of $\Delta f$ around frequency $f$ are in the box and the resistor is connected to the electrodes of the antenna. We start from thermal equilibrium, *i.e.*, a uniform temperature within the box, including the walls, the inherent thermal radiation, the antenna, the resistor, and the thermal isolation/screening.

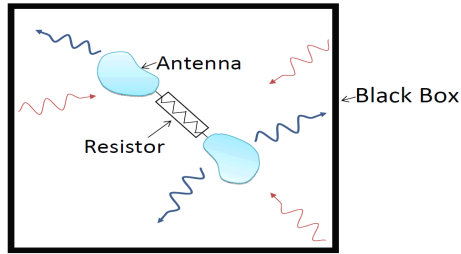

Figure 1. Outline of the gedanken experiment

Conditions *i*) and *ii*) will result in the following situation:

*a*) In accordance with condition *i*), the energy supplied by the resistor will be radiated by the antenna in the *M*-th orbital wave mode. This energy will be absorbed in the walls. The wall will emit thermal radiation in the form of plane waves [10] with a power given by Eq. 4.

*b*) In accordance with condition *ii*), the antenna can pick up a signal only at the *M*-th orbital mode while it will discard all plane wave components radiated by the walls of the box. This means that the antenna will not pick up any signal because the walls emit only plane waves [10].

Thus the energy will flow out from the resistor and cannot return. Therefore Boltzmann's Principle of Detailed Balance [12] cannot be satisfied. The resistor cools down, which implies that a temperature inhomogeneity is induced in the system in thermal equilibrium and hence the Second Law of Thermodynamics is violated. The only way to avoid violation of the Second Law of Thermodynamics with the above set-up is to suppose that the antenna also picks up plane wave modes. However, in that case the antenna cannot offer a separate information channel for the orbital mode.

It should also be mentioned that it is well-known that corresponding antenna types that can emit circularly polarized waves (which also have non-zero angular momentum) are sensitive to plane waves because a plane wave will excite its relevant polarization mode. Thus a circularly polarized antenna will not violate the Second Law when it is used in the same gedanken experiment as described above.

## Methods and Conclusions

We have presented and analyzed a gedanken experiment with a black body and a twisted-wave antenna in thermal equilibrium. We have shown that the assumption that at a single frequency more than two independent information channels can be provided by an antenna violates the Second Law of Thermodynamics. In conclusion, twisted waves cannot carry information that is independent from the information contained in plane wave modes at the same frequency.

## Acknowledgments

Discussions with Mihaly Benedict are appreciated. We are grateful for discussions and constructive comments to Claes-Göran Granqvist, Carl-Gustaf Ribbing, Derek Abbott, Ove Edfors, Julien Perruisseau-Carrier, Kai Chang and Greg Huff. We are grateful to the Authors of [5] and [6] for contacting us due to our manuscript in vixra.org and sending us their papers of high relevance; paper [5]: Ove Edfors and Anders Johansson; and paper [6]: Julien Perruisseau-Carrier, Michele Tamagnone and Christophe Craeye. We also grateful for email discussions with Bo Thide.